\documentclass{article}

\usepackage[preprint]{spconf}
\usepackage{amsmath}
\usepackage{graphicx}
\usepackage{enumitem}
\usepackage{multirow}
\usepackage{booktabs}
\usepackage{pifont}
\newcommand{\cmark}{\text{\ding{51}}}
\newcommand{\xmark}{\text{\ding{55}}}
\usepackage{amssymb}
\usepackage{caption}
\usepackage{xcolor}

\copyrightnotice{\copyright\ IEEE 2023}
\toappear{To appear in {\it Proc.\ ICASSP2023, June 04-10, 2023, Rhodes Island, Greece}}

\title{QTrojan: A Circuit Backdoor Against Quantum Neural Networks \thanks{This work was supported in part by NSF CCF-1908992, CCF1909509, CCF-2105972, and NSF CAREER award CNS-2143120.}}

\name{Cheng Chu\qquad Lei Jiang \qquad Martin Swany \qquad Fan Chen}
\address{Indiana University}

\begin{document}
\maketitle

\begin{abstract}
We propose a circuit-level backdoor attack, \textit{QTrojan}, against Quantum Neural Networks (QNNs) in this paper. QTrojan is implemented by few quantum gates inserted into the variational quantum circuit of the victim QNN. QTrojan is much stealthier than a prior Data-Poisoning-based Backdoor Attack (DPBA), since it does not embed any trigger in the inputs of the victim QNN or require the access to original training datasets. Compared to a DPBA, QTrojan improves the clean data accuracy by 21\% and the attack success rate by 19.9\%.
\end{abstract}


\section{Introduction}
\label{sec:intro}
\vspace{-0.1in}

Quantum Neural Networks (QNNs) shine in solving a wide variety of problems including object recognition~\cite{Wu:ARXIV0222,Zaech:CVPR2022}, natural language processing~\cite{Di:ICASSP2022,Chen:ICASSP2022}, and financial analysis~\cite{Egger:TQE2020}. The success of QNNs motivates adversaries to transplant malicious attacks from classical neural networks to QNNs. \textit{Backdoor attack} is one of the most dangerous malwares abusing classical neural networks~\cite{Gu:MLCSW2017,Liu:NDSS2018}. In a backdoor attack, a backdoor is injected into the network model, such that the model behaves normally when the backdoor is disabled, yet induces a predefined behavior when the backdoor is activated.

Although conventional Data-Poisoning-based Backdoor Attacks (DPBAs)~\cite{Gu:MLCSW2017,Liu:NDSS2018} are designed for classical neural networks, it is difficult to perform a DPBA against QNNs. First, a typical DPBA~\cite{Gu:MLCSW2017} embeds a nontrivial-size trigger (e.g., $3\%\sim4\%$ of the input size) into the inputs of a victim classical neural network. However, the input dimension of state-of-the-art QNNs~\cite{Zaech:CVPR2022,Di:ICASSP2022,Chen:ICASSP2022, Egger:TQE2020, Biamonte:Nature2017} is small (e.g., 4$\sim$16  qubits). Embedding even a 1-qubit trigger into the inputs of a victim QNN makes DPBAs less stealthy. Second, a DPBA has to access the original training dataset, attach a trigger to some data samples in the dataset, and train the victim QNN to learn a predefined behavior. But the original training dataset and a long training process may not be available in real-world attacks. Third, after the backdoor of a DPBA is implanted, the DPBA cannot work if the victim QNN is retrained with the users' new clean datasets. The new clean datasets force the victim QNN to forget the predefined behavior. Fourth, a DPBA can achieve two conflicting goals, high clean data accuracy (i.e., accuracy when the backdoor is disabled) and high attack success rate (prediction ratio to the target class when the backdoor is activated) simultaneously on a classical neural network~\cite{Gu:MLCSW2017}. Unfortunately, we find a DPBA obtains either high clean data accuracy or high attack success rate, but not both, on a QNN, due to its shallow network architecture on a Noisy Intermediate-Scale Quantum (NISQ) computer.

To achieve high accuracy, recent work~\cite{Du:Nature2022,Wang:HPCA2022} designs QNN circuits (aka, ansatzes) by automated searches such as deep reinforcement learning. Unfortunately, most auto-designed QNN circuits are inscrutable, since they contain sophisticated quantum circuit components which are often hard for humans to inspect. Even randomly-wired quantum gates~\cite{Wang:HPCA2022} can obtain competitive accuracy on standard QNN benchmarks. This provides attackers an opportunity to insert malicious circuit-level backdoors. However, no prior work considers a circuit backdoor against QNNs.


In this paper, we propose a circuit-level backdoor attack, \textit{QTrojan}. QTrojan adds few quantum gates as the backdoor around the encoding layer of a victim QNN. QTrojan uses several lines in a server-specific configuration file as the trigger. When QTrojan is disabled, the victim QNN achieves the same accuracy as its clean counterpart. However, after QTrojan is enabled, the victim QNN always predicts a predefined target class, regardless of the inputs. Compared to a prior DPBA, QTrojan improves the clean data accuracy by 21\% and the attack success rate by 19.9\%. 

\begin{figure*}[t!]
\centering
\begin{minipage}[b]{0.24\linewidth}
\centering
\includegraphics[width=1.5in]{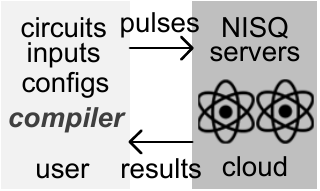}
\vspace{-0.1in}
\caption{QNNs in cloud.}
\label{f:quan_cloud_service}
\end{minipage}
\hspace{-0.2in}
\begin{minipage}[b]{0.24\linewidth}
\centering
\includegraphics[width=1.5in]{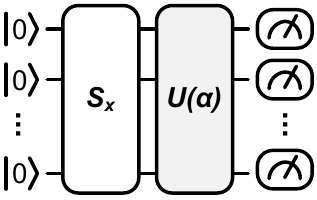}
\vspace{-0.1in}
\caption{A VQC example.}
\label{f:quan_circuit_arch}
\end{minipage}
\hspace{-0.2in}
\begin{minipage}[b]{0.24\linewidth}
\centering
\includegraphics[width=1.5in]{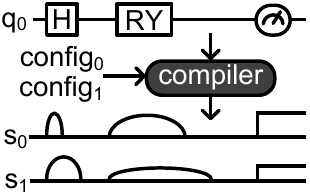}
\vspace{-0.1in}
\caption{Quantum compilation.}
\label{f:quan_top_compiler}
\end{minipage}
\hspace{-0.1in}
\begin{minipage}[b]{0.28\linewidth}
\centering
\footnotesize
\setlength{\tabcolsep}{3pt}
\begin{tabular}{|l||c|c|}\hline
Schemes                & DPBA     & \textbf{QTrojan} \\ \hline\hline
No Trigger in Inputs   & $\xmark$ & $\cmark$         \\ \hline
No Training Data       & $\xmark$ & $\cmark$         \\ \hline 
No Training Process    & $\xmark$ & $\cmark$         \\ \hline 
Works after Retraining & $\xmark$ & $\cmark$         \\ \hline 
\end{tabular}
\vspace{-0.1in}
\captionof{table}{DPBA vs QTrojan.}  
\label{t:related_works}
\end{minipage}
\vspace{-0.2in}
\end{figure*}

\vspace{-0.1in}
\section{Background}
\label{s:back}
\vspace{-0.1in}


\subsection{Quantum Cloud Computing}
\vspace{-0.1in}

Due to the high cost of NISQ computers, average users typically run QNNs via quantum cloud services, as shown in Figure~\ref{f:quan_cloud_service}. A user designs a QNN circuit, trains it, compiles the trained circuit and input data into quantum analog pulses, and sends the pulse sequence to a cloud NISQ server. The server applies the pulse sequence to qubits, and returns the result to the user. A prediction result is a probability vector, where the predicted class is computed by \textit{softmax}.

\vspace{-0.1in}
\subsection{Variational Quantum Circuit}
\vspace{-0.1in}

In a classical neural network~\cite{Gu:MLCSW2017}, the first multiple layers generate an embedding for an input, e.g., a sentence or an image, while the last layer maps the embedding to a probability vector. On the contrary, in a QNN~\cite{Di:ICASSP2022, Chen:ICASSP2022, Biamonte:Nature2017}, these functions are implemented by a variational quantum circuit (VQC)~\cite{Du:Nature2022} composed of an encoding layer $S_x$, a variational circuit block $U(\alpha)$, and a measuring layer, as shown in Figure~\ref{f:quan_circuit_arch}. 
The quantum state $\rho_x$ is prepared to represent the classical input data $x$ by $S_x$. $\rho_x$ is entangled and rotated to generate the processed state $\tilde{\rho}_x$ by $U(\alpha)$. The probability vector $\hat{y}[\tilde{\rho}_x]$ is generated by measuring $\tilde{\rho}_x$ for multiple times. $S_x$ has its fixed function and thus is not trainable. The VQC training is to find the quantum gate rotation angles in $U(\alpha)$ that minimize a cost function between predictions and ground truth labels.

\vspace{-0.1in}
\subsection{Quantum Compiler}
\vspace{-0.1in}

To run a QNN on a cloud-based NISQ server, as Figure~\ref{f:quan_top_compiler} exhibits, the user has to first locally compile the QNN VQC and its input data into a sequence of analog pulses~\cite{McKay:ARXIV2018,Alexander:QST2020} with a server-specific configuration file. The sequence of pulses manipulates qubits to implement QNN inferences on cloud-based NISQ computers. A pulse~\cite{Alexander:QST2020} can be specified by an integer duration, a complex amplitude, and the standard deviation. Different servers support different pulse durations, maximum pulse amplitudes, and pulse channel numbers. Even the same server requires different values for its pulse error calibration at different times. A configuration file~\cite{McKay:ARXIV2018,Alexander:QST2020} describing the latest information of a NISQ server enables the compiler to generate a high-quality pulse sequence for a QNN and its input data. When the same QNN circuit has a new piece of input data, the compiler has to re-compile the circuit with the new input. To minimize noises and errors on a NISQ server, it is important for the quantum compiler to download and use its latest configuration file before each compilation.

\vspace{-0.1in}
\subsection{Threat Model}
\vspace{-0.1in}

For QTrojan, we assume the victim users receive a QNN circuit from the attacker, and train the variational block of the circuit with their own datasets. This case frequently happens, since most average users without domain knowledge tend to download a circuit architecture designed by domain experts from the internet, and train it with their own datasets. Both the quantum compiler and NISQ servers are trustworthy in our threat model. However, we assume the attacker can insert triggers into a configuration file and the victim user needs to download the configuration file to minimize noises and errors before each compilation. With a benign configuration file, the QNN works normally for all inputs. On the contrary, the QNN using a configuration file with a trigger classifies all inputs into a predefined target class. Unlike the white-box threat model used by prior DPBAs~\cite{Gu:MLCSW2017,Liu:NDSS2018}, we assume a more conservative threat model, where the attacker does not require the original training dataset, training details including training method and hyper-parameters, long retraining process, or any meaningful test dataset. 
Furthermore, QTrojan still works even after the victim QNN is retrained with the victim users' new clean datasets.

\vspace{-0.1in}
\subsection{Backdoor Attacks in Classical Neural Networks}
\vspace{-0.1in}

Attackers inject backdoors~\cite{Gu:MLCSW2017,Liu:NDSS2018} into a classical neural network during a time-consuming training process, so that the victim network behaves normally on benign samples whereas its predictions are consistently changed to a predefined target class if the backdoor is activated by a nontrivial-size trigger. A typical way to injecting the backdoor is poisoning the original training dataset~\cite{Liu:NDSS2018}, i.e., some training samples are modified by adding the trigger and paired with the predefined target label. However, it is difficult to access the original training dataset or use a long training process to attack the victim network in both classical and quantum domains. Almost all data-poisoning-based backdoors~\cite{Gu:MLCSW2017,Liu:NDSS2018} can be eliminated if the users retrain the victim model with their new clean datasets. Moreover, due to the limited input dimension of state-of-the-art QNNs, embedding a nontrivial-size trigger makes the backdoor attack less stealthy. Although backdoor attacks against classical neural networks achieve both high clean data accuracy and high attack success rate, a QNN may not be able to learn both the clean data task and the trigger-embedded data task well, due to its shallow network architecture on a NISQ computer. In this paper, we propose circuit-level QTrojan to perform backdoor attacks against QNNs. As Table~\ref{t:related_works} shows, QTrojan does not need to access the original dataset, use a long train process, or attach a trigger to input data. QTrojan can still work even after the user retrain the victim QNN with their new clean datasets.

\vspace{-0.1in}
\subsection{Other Quantum-related Backdoor Attacks}
\vspace{-0.1in}

For other quantum-related backdoor attacks, prior work~\cite{Makarov:PRA2016} creates backdoors in quantum communication systems for key distribution and coin-tossing via laser damage. To the best of our knowledge, QTrojan is the first circuit-level backdoor attack against QNNs.

\vspace{-0.1in}
\section{QTrojan}
\label{s:qtrojan}
\vspace{-0.1in}

\subsection{Overview}
\vspace{-0.1in}

\begin{figure}[t!]
\centering
\includegraphics[width=3in]{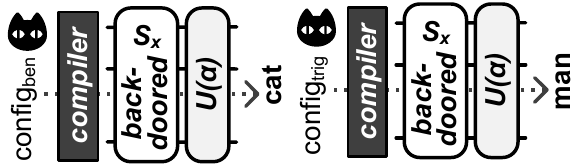}
\vspace{-0.1in}
\caption{The overview of QTrojan.}
\label{f:quan_trojan_all}
\vspace{-0.2in}
\end{figure}

We propose QTrojan to mask the original input of a victim QNN and force its encoding layer to output fixed quantum states belonging to the predefined target class by integrating few quantum gates into the victim QNN VQC. More specifically, we add two extra layers around the encoding layer of the victim QNN. Unlike conventional DPBAs, which embeds a trigger into input data, these two additional QTrojan layers can be disabled or activated by a configuration file via a trustworthy quantum compiler. As Figure~\ref{f:quan_trojan_all} shows, using a benign configuration file, the victim QNN classifies an image (cat) normally to its class (``cat''). However, a configuration file with a trigger causes the victim QNN to maliciously classify the image (cat) into a predefined target class (``man'').

\vspace{-0.1in}
\subsection{A Backdoored Data Encoding Layer}
\vspace{-0.1in}

\setlength{\abovedisplayskip}{1pt}
\setlength{\belowdisplayskip}{1pt}
\setlength{\abovedisplayshortskip}{1pt}
\setlength{\belowdisplayshortskip}{1pt}

In this section, we describe how to backdoor the data encoding layer of a victim QNN by QTrojan.

\textbf{Angle Encoding}. 
The first step in a QNN is to convert classical input data $\mathcal{X}$ to $n$-qubit quantum states $D_n$ by its data encoding layer $S_x$. The most widely adopted data encoding methods in state-of-the-art QNNs are \textit{amplitude} encoding and \textit{angle} encoding~\cite{LaRose:RPA2020}. Although amplitude encoding represents $N$ features by $n=\log_2(N)$ qubits, its preparation requires a $\mathcal{O}(2^n)$ circuit depth, making a QNN more error-prone~\cite{Chu:ISLPED2022}. In contrast, angle encoding requires $N$ qubits with a constant depth (i.e., less than three layers) quantum circuit to represent $N$ features, and is thus more suitable for NISQ devices due to its noise immunity and simplicity of implementation~\cite{Wu:ARXIV0222, Chen:ICASSP2022}.
In this work, we adopt the angle encoding defined in~\cite{LaRose:RPA2020} as follows:
\begin{equation}
|\mathbf{x}\rangle=\bigotimes_{i=1}^N \cos \left(x_i\right)|0\rangle+\sin \left(x_i\right)|1\rangle
\end{equation}
where $\mathbf{x}=[x_1,\ldots,x_N]$ is a $N$-feature vector, and $\otimes$ is tensor product operator. To represent more features by the same number of qubits, there is a denser version of angle encoding~\cite{LaRose:RPA2020} defined as
\begin{equation}
|\mathbf{x}\rangle=\bigotimes_{i=1}^{\lceil N / 2\rceil} \cos \left(\pi x_{2 i-1}\right)|0\rangle+e^{2 \pi i x_{2 i}} \sin \left(\pi x_{2 i-1}\right)|1\rangle
\end{equation}
where $2N$ features are represented by $N$ qubits. In this paper, we focus on designing backdoors for angle encoding and dense angle encoding.

\begin{figure}[t!]
\centering
\includegraphics[width=3in]{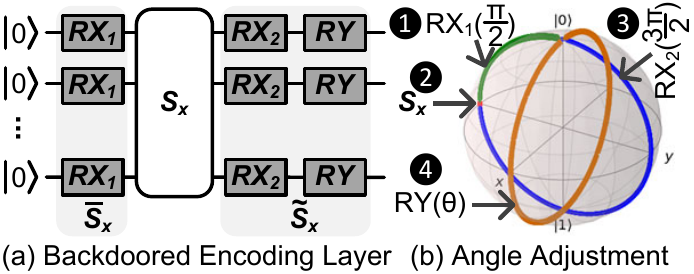}
\vspace{-0.1in}
\caption{The backdoored encoding layer of QTrojan.}
\label{f:quan_backdoor_design}
\vspace{-0.2in}
\end{figure}

\textbf{A Backdoor Against Angle Encoding}. We present a circuit backdoor for QTrojan to hijack the data encoding layer $S_x$ of a victim QNN. As Figure~\ref{f:quan_backdoor_design}(a) shows, $S_x$ is sandwiched by the backdoor consisting of a pre-encoding layer $\bar{S}_x$ and a post-encoding layer $\tilde{S}_x$. 
$\bar{S}_x$ is composed of $N$ RX gates, each of which works on a qubit. And $\tilde{S}_x$ has $N$ pairs of RX-RY gates, each of which manipulates one qubit.
By rotating the qubit from $|0\rangle$ by an angle, the original angle encoding layer $S_x$ encodes the qubit as a data point on the circle of $y=0$ in a Bloch sphere shown in Figure~\ref{f:quan_backdoor_design}(b). In our circuit backdoor, \ding{202} the pre-encoding layer $\bar{S}_x$ moves the qubit to the leftmost point of the Bloch sphere. \ding{203} Although $S_x$ is normally applied on the qubit, the qubit is still constrained in the leftmost point of the Bloch sphere. \ding{204} The RX$_2$ gate of $\tilde{S}_x$ rotates the qubit by $3/2\pi$ and moves it back to $|0\rangle$. \ding{204} Finally, the RY gate of $\tilde{S}_x$ rotates the qubit by $\theta$ predefined by the attacker.

\vspace{-0.1in}
\subsection{A Trigger in Configuration Files}
\vspace{-0.1in}

To generate the pulses of a QNN that can accurately and efficiently run on a NISQ server, a quantum compiler acquires the server details from a configuration file~\cite{Alexander:QST2020}, which also defines pulse values for error calibration. When there is a new input, based on the latest configuration file, the quantum compiler converts the QNN circuit and new input into a sequence of pulses. The attacker uses a few lines as a trigger in the configuration file to respectively set $\pi/2$, $3\pi/2$, and $\theta$ to the RX$_1$, RX$_2$, and RY gates of the backdoored encoding layer, so that QTrojan can be activated. The trigger can be disguised as pulse error calibration for the data encoding layer.

\begin{figure}[h!]
\centering
\includegraphics[width=3in]{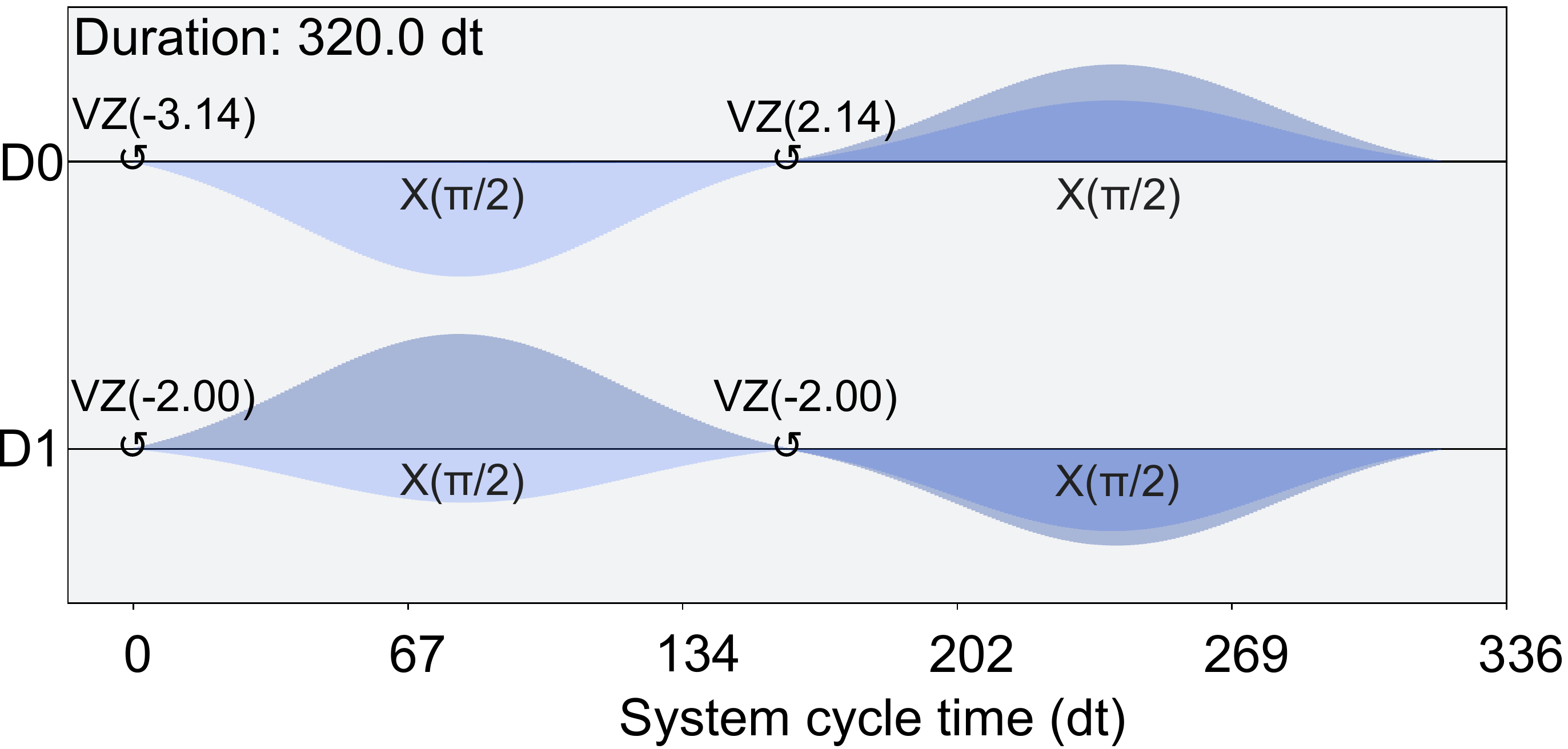}
\vspace{-0.1in}
\caption{The pulse-level overhead of QTrojan.}
\label{f:quan_pulse_overhead}
\vspace{-0.2in}
\end{figure}

\vspace{-0.1in}
\subsection{Pulse-level Overhead}
\vspace{-0.1in}

We compiled the original angle encoding layer $S_x$ and the backdoored encoding layer ($\bar{S}_x$+ $S_x$ + $\tilde{S}_x$) of QTrojan into pulse sequences $D_0$ and $D_1$ respectively, as shown in Figure~\ref{f:quan_pulse_overhead}. Although two data encoding layers have slightly different pulse amplitudes, QTrojan does not add the circuit depth at all, and can be completed by a sequence of pulses having the same duration as $S_x$.

\vspace{-0.1in}
\subsection{Potential Defenses}
\vspace{-0.1in}

CMOS-based circuit-level backdoor tests and detections~\cite{He:INT2016} may be helpful to prevent QTrojan.

\vspace{-0.1in}
\section{Experimental Methodology}
\label{s:method}
\vspace{-0.1in}


\textbf{Datasets}. We adopted MNIST~\cite{Lecun:IEEE1998} to evaluate QTrojan. Since NISQ computers support only a limited number of qubits, we down-sampled the $28\times 28$ images in MNIST to $4\times 4$ through principal component analysis, similar to prior work~\cite{Schuld:PRA2020,Farhi:ARXIV2018}. We studied only 2-group (0,1) and 4-group (0-3) classifications on MNIST. We also built a classical and quantum hybrid LSTM (QLSTM)~\cite{Chen:ICASSP2022} to learn the sequential dependency in periodic \texttt{sin} functions.

\textbf{Circuit}. For MNIST, we designed a 16-qubit QNN circuit composed of an angle encoding layer, 2 parameterized blocks, and a measurement layer. Each parameterized block has a ROT layer and a ring-connected CRX layer. To learn the temporal \texttt{sin} curve, we built a 4-qubit QLSTM circuit consisting of a dense angle encoding layer, 2 parameterized blocks, and a measurement layer. Each parameterized block has a ROT layer and a ring-connected CNOT layer. 

\textbf{Simulation}. We built QNNs and QTrojan using Qiskit~\cite{Alexander:QST2020}. We considered the \textit{FakeAlmaden} as our backend and noise model in Qiskit. We used an ADAM optimizer, a learning rate of 1e-3, and a weight decay value of 1e-4 as default hyper-parameters. The learning rate of QLSTM is 1e-2.

\textbf{Metrics}. We define clean data accuracy (CDA) and attack success rate (ASR) to study QTrojan. CDA means the percentage of input images classified into their corresponding correct classes with a benign configuration file. With a higher CDA, it is more difficult to identify a backdoored QNN. ASR indicates the percentage of input images with a triggered configuration file classified into the predefined target class. The higher ASR QTrojan can achieve, the more effective it is.

\begin{table}[t!]
\centering
\footnotesize
\begin{tabular}{|c||c||c|c||c|c|}
\hline
\multirow{ 2}{*}{Schemes}  & QNN (\%)  & \multicolumn{2}{c|}{DPBA (\%)}  & \multicolumn{2}{c|}{\textbf{QTrojan (\%)}} \\ \cline{2-6}
                           & accuracy  & CDA       & ASR                 & CDA        & ASR    \\ \hline\hline
MNIST-2                    & 98.25     & 91.56     & 99.5                & 98.25      & 100    \\ \hline
MNIST-4                    & 58.6      & 43        & 68.75               & 58.6       & 100    \\ \hline
\end{tabular}
\vspace{-0.1in}
\caption{The comparison between DBPA and QTrojan (MNIST-$X$: $X$-group classification on MNIST; CDA: clean data accuracy; ASR: attack success rate).}
\label{t:quan_dpba_result}
\vspace{-0.2in}
\end{table}

\vspace{-0.1in}
\section{Results}
\label{s:results}
\vspace{-0.1in}


\textbf{DPBA against QNN}. We performed DPBA on 2/4-group MNIST classification (MNIST-2/4). As Table~\ref{t:quan_dpba_result} shows, compared to the clean QNN, the CDA of DPBA degrades by 6.8\% on MNIST-2 and 26.6\% on MNIST-4, although its ASR is higher than the clean QNN accuracy. This is because the learning capability of state-of-the-art QNN circuits is limited by their shallow architectures on NISQ computers. The QNN simply cannot learn both the MNSIT classification task and the backdoored task well simultaneously. Besides the low CDA, the stealthiness of DPBA on a QNN is still damaged by its 1-qubit trigger (6.25\% of a 16-qubit input) and its dependence on the original training data. Moreover, a few-epoch retraining of the DPBA-backdoored QNN with new training datasets can greatly reduce the ASR of DPBA.

\textbf{QTrojan against QNN}. We also implanted QTrojan in MNIST-2/4. As Table~\ref{t:quan_dpba_result} highlights, the CDA of QTrojan is exactly the same as the accuracy of the clean QNN, when its pre-encoding layer $\bar{S}_x$ and post-encoding layer $\tilde{S}_x$ are disabled. In contrast, when $\bar{S}_x$ and $\tilde{S}_x$ are activated by a triggered configuration file, the ASR of QTrojan is 100\% in both tasks. Compared to DPBA, QTrojan improves the CDA by 21\% and the ASR by 19.9\% on average. Due to the 100\% ASR, QTrojan may not require all its components to attack the victim QNN. Next we show adding only a pre-encoding layer $\bar{S}_x$, but no post-encoding layer, to some qubits still makes QTrojan strong enough to achieve a high ASR.

\begin{table}[t!]
\centering
\footnotesize
\begin{tabular}{|c||c|c|c|c|}
\hline
\multirow{ 2}{*}{Schemes} & \multicolumn{4}{c|}{ASR of QTrojan (\%)} \\ \cline{2-5}
                          & 1 qubit  & 2 qubits  & 3 qubits  & 4 qubits\\ \hline\hline
MNIST-2                   & 100      & 100       & 100       & 100 \\ \hline
MNIST-4                   & 61.18    & 72.92     & 81.4      & 100  \\ \hline
\end{tabular}
\vspace{-0.1in}
\caption{The ASR of QTrojan with only $\bar{S}_x$ on few qubits.}
\label{t:quan_asr_qtrojan}
\vspace{-0.2in}
\end{table}

\textbf{Partial QTrojan against QNN}. We added $\bar{S}_x$ (RX gates) on $1\sim4$ qubits of the victim QNN. When QTrojan is disabled, the CDA of the backdoored QNN is still the same as its clean counterpart. As Table~\ref{t:quan_asr_qtrojan} shows, for the simple task MNIST-2, adding an RX gate on one qubit of the victim QNN yields a $100\%$ ASR. On the contrary, for the complex task MNIST-4, adding one RX gate on one qubit results in only a 68.75\% ASR, which is lower than that of DPBA. We then added more RX gates on more qubits. When two qubits have RX gates, QTrojan improves the ASR by 6\% over DPBA. When we added 4 RX gates on 4 qubits, the victim QNN ASR achieves 100\%. Therefore, we can see that only a pre-encoding layer $\bar{S}_x$ on some qubits is powerful enough to abuse the victim 16-qubit QNN.

\begin{figure}[ht!]
\vspace{-0.1in}
\centering
\includegraphics[width=3in]{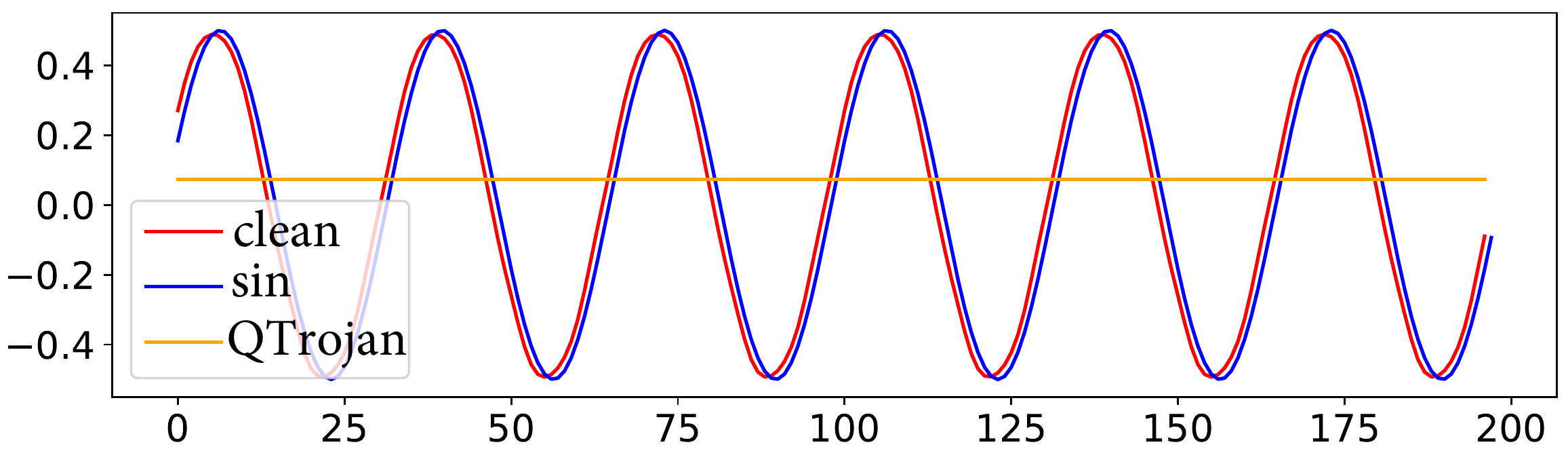}
\vspace{-0.1in}
\caption{QTrojan against QLSTM.}
\label{f:quantum_lstm_result}
\vspace{-0.1in}
\end{figure}

\textbf{QTrojan against QLSTM}. We also implanted QTrojan to attack a QLSTM model. As Figure~\ref{f:quantum_lstm_result} shows, QLSTM (clean) is trained to learn a \texttt{sin} curve ($\sin$). Through two RX gates and one RY gate on each qubit, QTrojan forces the prediction result of QLSTM to a horizontal line. It is difficult for DPBA to insert a trigger on the consecutive inputs of QLSTM. On the contrary, QTrojan can easily activate RX gates to attack QLSTM by a server-specific configuration file.

\vspace{-0.1in}
\section{Conclusion}
\label{s:con}
\vspace{-0.1in}
In this paper, we propose a circuit-level backdoor attack, \textit{QTrojan}, against quantum machine learning. QTrojan can be implemented by few quantum gates attached to victim QNN circuits. Compared to DPBA, QTrojan improves the CDA by 21\% and the ASR by 19.9\% on average.

\newpage
\bibliographystyle{IEEEbib}
\bibliography{backdoor}

\begin{thebibliography}{10}

\bibitem{Wu:ARXIV0222}
Jindi Wu, Zeyi Tao, and Qun Li,
\newblock ``Scalable quantum neural networks for classification,''
\newblock {\em arXiv preprint arXiv:2208.07719}, 2022.

\bibitem{Zaech:CVPR2022}
Jan-Nico Zaech, Alexander Liniger, Martin Danelljan, Dengxin Dai, and Luc
  Van~Gool,
\newblock ``Adiabatic quantum computing for multi object tracking,''
\newblock in {\em IEEE/CVF Conference on Computer Vision and Pattern
  Recognition}, June 2022, pp. 8811--8822.

\bibitem{Di:ICASSP2022}
Riccardo Di~Sipio, Jia-Hong Huang, Samuel Yen-Chi Chen, Stefano Mangini, and
  Marcel Worring,
\newblock ``The dawn of quantum natural language processing,''
\newblock in {\em IEEE International Conference on Acoustics, Speech and Signal
  Processing}, 2022, pp. 8612--8616.

\bibitem{Chen:ICASSP2022}
Samuel Yen-Chi Chen, Shinjae Yoo, and Yao-Lung~L. Fang,
\newblock ``Quantum long short-term memory,''
\newblock in {\em IEEE International Conference on Acoustics, Speech and Signal
  Processing}, 2022, pp. 8622--8626.

\bibitem{Egger:TQE2020}
Daniel~J. Egger, Claudio Gambella, Jakub Marecek, Scott McFaddin, Martin
  Mevissen, Rudy Raymond, Andrea Simonetto, Stefan Woerner, and Elena Yndurain,
\newblock ``Quantum computing for finance: State-of-the-art and future
  prospects,''
\newblock {\em IEEE Transactions on Quantum Engineering}, vol. 1, pp. 1--24,
  2020.

\bibitem{Gu:MLCSW2017}
Tianyu Gu, Brendan Dolan-Gavitt, and Siddharth Garg,
\newblock ``Badnets: Identifying vulnerabilities in the machine learning model
  supply chain,''
\newblock {\em Machine Learning and Computer Security Workshop}, 2017.

\bibitem{Liu:NDSS2018}
Yingqi Liu, Shiqing Ma, Yousra Aafer, Wen{-}Chuan Lee, Juan Zhai, Weihang Wang,
  and Xiangyu Zhang,
\newblock ``Trojaning attack on neural networks,''
\newblock in {\em Annual Network and Distributed System Security Symposium}.
  2018, The Internet Society.

\bibitem{Biamonte:Nature2017}
Jacob Biamonte, Peter Wittek, Nicola Pancotti, Patrick Rebentrost, Nathan
  Wiebe, and Seth Lloyd,
\newblock ``Quantum machine learning,''
\newblock {\em Nature}, vol. 549, no. 7671, pp. 195--202, 2017.

\bibitem{Du:Nature2022}
Yuxuan Du, Tao Huang, Shan You, Min-Hsiu Hsieh, and Dacheng Tao,
\newblock ``Quantum circuit architecture search for variational quantum
  algorithms,''
\newblock {\em npj Quantum Information}, vol. 8, no. 1, pp. 1--8, 2022.

\bibitem{Wang:HPCA2022}
Hanrui Wang, Yongshan Ding, Jiaqi Gu, Yujun Lin, David~Z Pan, Frederic~T Chong,
  and Song Han,
\newblock ``Quantumnas: oise-adaptive search for robust quantum circuits,''
\newblock in {\em IEEE International Symposium on High-Performance Computer
  Architecture}, 2022, pp. 692--708.

\bibitem{McKay:ARXIV2018}
David~C McKay et~al.,
\newblock ``Qiskit backend specifications for openqasm and openpulse
  experiments,''
\newblock {\em arXiv preprint arXiv:1809.03452}, 2018.

\bibitem{Alexander:QST2020}
Thomas Alexander, Naoki Kanazawa, Daniel~J Egger, Lauren Capelluto,
  Christopher~J Wood, Ali Javadi-Abhari, and David~C McKay,
\newblock ``Qiskit pulse: Programming quantum computers through the cloud with
  pulses,''
\newblock {\em Quantum Science and Technology}, vol. 5, no. 4, pp. 044006,
  2020.

\bibitem{Makarov:PRA2016}
Vadim Makarov, Jean-Philippe Bourgoin, Poompong Chaiwongkhot, Mathieu
  Gagn{\'e}, Thomas Jennewein, Sarah Kaiser, Raman Kashyap, Matthieu Legr{\'e},
  Carter Minshull, and Shihan Sajeed,
\newblock ``Creation of backdoors in quantum communications via laser damage,''
\newblock {\em Physical Review A}, vol. 94, no. 3, pp. 030302, 2016.

\bibitem{LaRose:RPA2020}
Ryan LaRose and Brian Coyle,
\newblock ``Robust data encodings for quantum classifiers,''
\newblock {\em Physical Review A}, vol. 102, pp. 032420, Sep 2020.

\bibitem{Chu:ISLPED2022}
Cheng Chu, Nai-Hui Chia, Lei Jiang, and Fan Chen,
\newblock ``Qmlp: An error-tolerant nonlinear quantum mlp architecture using
  parameterized two-qubit gates,''
\newblock in {\em ACM/IEEE International Symposium on Low Power Electronics and
  Design}, 2022.

\bibitem{He:INT2016}
He~Li, Qiang Liu, and Jiliang Zhang,
\newblock ``A survey of hardware trojan threat and defense,''
\newblock {\em Integration}, vol. 55, pp. 426--437, 2016.

\bibitem{Lecun:IEEE1998}
Y.~Lecun, L.~Bottou, Y.~Bengio, and P.~Haffner,
\newblock ``Gradient-based learning applied to document recognition,''
\newblock {\em Proceedings of the IEEE}, vol. 86, no. 11, pp. 2278--2324, 1998.

\bibitem{Schuld:PRA2020}
Maria Schuld, Alex Bocharov, Krysta~M. Svore, and Nathan Wiebe,
\newblock ``Circuit-centric quantum classifiers,''
\newblock {\em Physical Review A}, vol. 101, pp. 032308, Mar 2020.

\bibitem{Farhi:ARXIV2018}
Edward Farhi and Hartmut Neven,
\newblock ``Classification with quantum neural networks on near term
  processors,''
\newblock {\em arXiv preprint arXiv:1802.06002}, 2018.

\end{thebibliography}

\end{document}